\documentclass[pre,twocolumn]{revtex4}

\usepackage{amsmath,amssymb}
\usepackage{graphicx}

\begin{document}

\title{Continuous Capillary Condensation}
\author{A.O.\ Parry$^1$, C.\ Rasc\'{o}n$^2$, N.B.\ Wilding$^3$ and R.\ Evans$^4$}
\affiliation{\vspace*{.25cm}\\
$^1$ Department of Mathematics, Imperial College London, London SW7 2AZ , United Kingdom\\
$^2$ GISC, Departamento de Matem\'{a}ticas, Universidad Carlos III de Madrid, 28911 Legan\'{e}s, Spain\\
$^3$ Department of Physics, University of Bath, Bath BA2 7AY, United Kingdom\\
$^4$ HH Wills Physics Laboratory, University of Bristol, Bristol BS8 1TL, United Kingdom }

\begin{abstract}
{\indent} We show that condensation in a capped capillary slit is a continuous interfacial critical phenomenon, related intimately to several other surface phase transitions. In three dimensions (3d), the adsorption and desorption branches correspond to the unbinding of the meniscus from the cap and opening, respectively and are equivalent to 2d-like complete-wetting transitions. For dispersion forces, the singularities on the two branches are distinct, owing to the different interplay of geometry and intermolecular forces. In 2d we establish precise connection, or covariance, with 2d critical-wetting and wedge-filling transitions, {\it i.e.} we establish that certain interfacial properties in very different geometries are identical. Our predictions of universal scaling and covariance in finite capillaries are supported by extensive Ising model simulation studies in 2d and 3d.
\end{abstract}

\maketitle

{\indent}Capillary condensation (CC) is central to our understanding of 
confined fluids and has received much attention over the last few 
decades \cite{RW}. As is well known, vapour confined in a slit of width $L$ condenses 
at a pressure $p_{co}(L)$, less than the value $p_{sat}$ at bulk saturation, given by the macroscopic {\it Kelvin equation } \cite{Thomson1871,Derjaguin1940},
\begin{equation}
 p_{sat}-p_{co}(L)=\frac{2\sigma \cos \theta}{L}+...
\label{Kelvin}
\end{equation}
where $\sigma$ is the liquid-vapour surface tension and $\theta$ the contact angle. Studies of CC, based on Landau or modern density functional methods, usually consider confining walls  of 
infinite area and assume translational 
invariance parallel to these \cite{RW, FNJCP83}. In this case, CC is certainly a first-order 
transition and mean-field treatments yield adsorption isotherms with a van der 
Waals loop. Far less attention has been given to CC in slits of 
finite depth $D$ which are capped at one end and open into a reservoir (see 
Fig. 1). This scenario is certainly experimentally accessible and is 
similar to recent analysis of adsorption on grooved and pitted substrates 
\cite{MisturaPRL02,OckoPRL06}. Three numerical studies
\cite{MarconivanSwolPRA1989,DarbellayYeomans93,TeksynDietrich06}, restricted to complete wetting ($\theta=0$), have reported the same basic finding: In a finite 
capillary, CC is a sharp but continuous transition (Fig. 1 ) and adsorption isotherms
 exhibit no van der Waals loops for any finite $D$ (see also \cite{Cohen,Nature}). However, the physical reason behind 
this striking change and the quantitative aspects of the transition have not been elucidated.

{\indent} Here we show that CC in such capped capillaries is a continuous interfacial 
critical phenomenon exhibiting observable critical singularities which are 
intimately related to several other surface phase transitions. Our two main
findings are: ({\bf{A}}) In 3d, adsorption and desorption in 
a deep capillary correspond to the continuous unbinding of the meniscus from 
the bottom and top, respectively, and map onto two-dimensional complete
 wetting \cite{Reviews,LipPRB85} with relevant scaling field $\Delta p = p_{co}(L)-p$. The
divergence of the average meniscus height on these branches is described by
the respective critical singularities $\langle\ell\rangle\sim\Delta 
p^{-\beta_A}$,$D-\langle\ell\rangle\sim|\Delta p|^{-\beta_D}$ which are, in general {\it{distinct}}. In particular, for 
dispersion forces, $\beta_A=\frac{1}{4}$ while $\beta_D=\frac{1}{3}$
implying that the adsorption (isotherm) is steeper than the desorption. For a 
finite-depth capillary, the meniscus has large-scale fluctuations at $p=p_{co}$ equivalent
to those of an interface in a 2d infinite capillary with {\it{opposing}}
walls for which there are long-standing predictions
\cite{ParryEvansNicolaidesPRL91,AlbanoBinderSurfSci1989,AbrahamSvrakicUptonPRL92,SteckiMaciolek}. ({\bf{B}}) In a 2d capped slit (or a 3d capillary pore), one can proceed further and relate CC to {\it {critical wetting}} transitions occurring at 2d planar substrates \cite{Reviews, AbrahamPRL80}. This precise connection is an extension of the geometrical 
covariance known for the filling of wedges
\cite{ParryGreenallWoodJPhysCond2002,AbrahamParryWoodEPL2002}, 
cones \cite{RasconParryPRL05} and apexes \cite{ParryGreenallJoseManuelPRL03} and implies that some (universal) 
interfacial properties in very different geometries are identical. These
predictions, based primarily on analysis of interfacial Hamiltonians, are supported fully by our Ising model simulation studies in 2d and 3d.

\begin{figure}[thb]
\hspace*{-.5cm}
\includegraphics[width=7.cm]{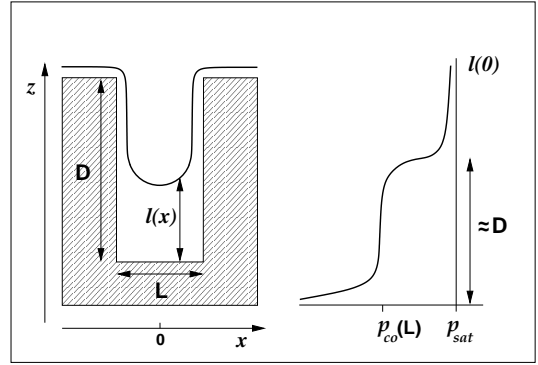}
\caption{Schematic cross-section of a capped capillary of depth $D$ and width $L$ illustrating the local interfacial height $\ell({\bf{x}})$. The slit is infinitely long in the $y$ direction. A typical adsorption isotherm is also sketched showing the rapid but continuous rise in the mid-point height near CC ($p\approx p_{co}(L)$) and the complete wetting as saturation is approached.
}\label{fig1}
\end{figure}

{\indent} In an infinite capillary, the two phases, referred to as capillary-liquid and 
capillary-vapour, coexist when $p=p_{co}(L)$. Any meniscus 
separating these phases is delocalised and has large-scale fluctuations, analogous to those of a planar interface between coexisting bulk phases, but in one lower dimension. Now in a capped system,
geometry necessitates the formation of a meniscus whose 
location is determined, in part, by the capillary thermodynamics. For 
$p< p_{co}(L)$ the meniscus must be located 
near the bottom because the capillary-liquid phase is metastable, while for 
$p>p_{co}(L)$, the capillary-vapour is metastable and the meniscus must 
reside near the capillary opening. The fundamental question is, what happens to the
meniscus as $p\to p_{co}$?

{\indent}We first follow \cite{DarbellayYeomans93} and introduce 
an interfacial model based on the local interfacial height 
$\ell(x,y)$, where $y$ is the variable along the capillary. The free energy is
\begin{equation}
H[\ell]=\sigma S[\ell] + \delta p\,V[\ell]+W[\ell]
\label{H1}
\end{equation}
Here $S$ is the fluid interfacial area, $V$ the volume of liquid, 
$\delta p=p_{sat}-p$, and $W$ is the binding potential accounting 
for the (dispersion) intermolecular forces. For confining 
walls which are completely wet ($\theta=0$), one can use 
\begin{equation}
W[\ell]\,=\,\frac{12\,A}{\pi}\,\int\!\!\!\!\int\! d{\bf r}\, d{\bf r}'\;\, \frac{1}{\;|{\bf r}-{\bf r}'|^6\,}
\label{W1}
\end{equation}
where $A>0$ is the Hamaker constant, an energy, and ${\bf r}$ and ${\bf r}'$ denote points 
in the substrate and vapour, respectively. The numerical prefactor is chosen so
that (\ref{W1}) recovers the usual binding potential per unit area
$W(\ell)=A\ell^{-2}$ for planar walls \cite{Reviews}. The model (\ref{H1}) describes the
whole adsorption isotherm including the CC and the complete wetting of the 
entire substrate as $p\rightarrow p_{sat}$. The latter is not of interest 
so we seek an effective Hamiltonian which 
will allow us to study the CC more easily. Accordingly we integrate 
out degrees of freedom keeping only the long-wavelength fluctuations in the 
meniscus height \textit{along} the capillary. Close to CC, the meniscus must be of 
semi-circular cross-section at local height $\ell(y)\equiv\ell(0,y)<D$. The
 fluctuations of the meniscus 
are then described by a {\it{capillary}} Hamiltonian, obtained via a constrained
 minimisation of (\ref{H1}). After some algebra we find \cite{tbp}
\begin{equation}
H_{cap}[\ell(y)]=\int\!dy\;\left\{\frac{\sigma L}{2}\left(\frac{d\ell}{dy}\right)^2 +
W_{cap}(\ell;D) \right\}
\label{H2}
\end{equation}
where the capillary binding potential is
\begin{equation}
W_{cap}(\ell;D)=\Delta p\,L\,\ell\;+\;\frac{5}{24}\frac{A\,L^2}{\ell^3}\;+\;\frac{A\,L}{(D-\ell)^2}\;+\,\cdots
\label{W2}
\end{equation}
 The model is only valid near CC and does not describe the complete wetting as 
$p\to p_{sat}$. Nevertheless it describes accurately the interplay of the 
geometry, forces and fluctuations for $p\approx p_{co}$, providing a
more transparent view of the interfacial behaviour.

{\indent} Eqn (\ref{W2}) is one of our central results and highlights some
intriguing properties of CC: ({\bf{I}}) The
 first term, conjugate to $\ell(y)$, is proportional to $\Delta p\equiv p_{co}(L)-p$, measuring the
 {\it{deviation from capillary coexistence}}. This establishes the connection
 with 
{\it{complete wetting}} phenomena, in one lower dimension, with the meniscus 
acting as the unbinding interface. ({\bf{II}}) Intermolecular forces repel the 
meniscus from the capillary cap and opening. At CC, the capillary cap wishes
 to be "wet" by the capillary-liquid while the capillary opening is "wet" by the
capillary-vapour. The finite-size (FS) effects mimic therefore those of an
interface in a capillary 
with {\it{opposing}} walls that are wet by different fluid phases 
\cite{ParryEvansPRL,Physica92,IndekeuEPL91,BinderLanFerrenPRL95}.
 This connection is deeper in 2d as we shall see later.
 ({\bf{III}}) While the repulsion from the top is similar to the planar wall \cite{Physica92},
that from the bottom is {\it{shorter-ranged}} due to a 
geometry-induced cancellation of intermolecular forces. This is the reason 
behind the 
asymmetry in the adsorption isotherm seen in numerical studies \cite{DarbellayYeomans93} and is a general
feature of CC in systems with long-ranged (dispersion) forces. For short-ranged forces the
(direct) repulsions from the top and bottom are similar 
to exponential decays.

{\indent}To continue, consider the critical behaviour in the
semi-infinite limit $D\to\infty$. The adsorption and desorption branches of 
the 3d capillary isotherm become analogous to 
2d-like complete wetting phase transitions describing the unbinding of
the meniscus from the cap and open end, respectively. The associated
critical singularities are well understood \cite{LipPRB85} and reflect the 
long-ranged forces
 presented in $W_{cap}(\ell)$ and fluctuation-effects associated with the 
wandering of the meniscus controlled by the surface tension term in
(\ref{H2}). The dependence on the slit width is significant since increasing $L$
effectively supresses the role of fluctuations. For example, the 
entropic repulsion from the bottom effectively adds a term 
$\propto(k_{B}T)^2/\sigma L\,\ell^2$ to (\ref{W2}). While, in principle, the
 asymptotic
divergence of $\ell$ is ultimately determined by 
this entropic repulsion, in practice it is irrelevant for slits
 more than a few angstroms in width because the amplitude is negligible
compared to that of the long-range forces. 
The latter dominate for all practical purposes and we anticipate
mean-field-like behaviour with  $\langle\,\ell\,\rangle\sim (A\, L /\Delta p)^{1/4}$, as quoted
earlier. On the 
desorption branch, fluctuation effects are similarly negligible, even though
they are marginal, and we predict  
$D-\langle\,\ell\,\rangle\sim (A/|\Delta p|)^{1/3}$. For systems with short-ranged forces, on the other
hand, the influence of fluctuations can no longer be neglected and both the 
adsorption
and desorption branches show the same critical behaviour. Thus, for example
on the adsorption branch $\langle\,\ell\,\rangle\sim (L^2\Delta p)^{-1/3}$. All these results 
are supported fully by
transfer-matrix analysis of the Hamiltonian (\ref{H2}).

{\indent}Condensation (or rather pseudo-condensation) in a 2d capped capillary
 is particularly interesting. Here we focus solely on systems with short-ranged 
forces since long-ranged forces are not pertinent for all physical scenarios. However, we broaden our discussion to the case of non-zero contact angle $\theta$. 
Firstly, as above, the adsorption and desorption in a semi-infinite capillary are related to the 1d
limit of complete wetting. The probability of finding the meniscus
at a given height is simply proportional to the Boltzmann weight of $W_{cap}$,
implying that on the adsorption branch $\langle\,\ell\,\rangle\sim (L\Delta p)^{-1}$ and
similarly for desorption. Thus the critical 
exponent $\beta_A=1$ can be identified as the $d\to 1^+$ limit of the general result
$\beta_{co}^{s}(d)=(3-d)/(d+1)$ appropriate to short-ranged complete wetting in dimension $d\le 3$ \cite{Reviews}.

 {\indent}To see the deep connection with 2d critical wetting we introduce what at
first appears to be an artificial geometry, and adopt a magnetic
notation also useful for comparison with our Ising
model simulation results. Consider a planar Ising-like system, with spontaneous
magnetisation $m_0$ and a boundary
(wall) in the shape of a trough. That is the height of the wall above some
reference line is $\Psi(x)=0$ for $\vert x\vert < L/2$ and $\Psi(x)=
(\vert x\vert-L/2)\tan\alpha $ otherwise. The spins on the boundary are subject 
to a surface field $h_1>0$. The spins away from the wall are subject to a 
position dependent external field $h(x)=h$ for $\vert x\vert < L/2$ 
and $h(x)=0^-$ otherwise. In general these boundary conditions induce a
fluctuating interface at height $\ell(x)$, which 
defines the "wetting" layer of up spins adsorbed near the wall.

 {\indent} The trough has limiting geometries, each having distinct
transitions. For $\alpha=\frac{\pi}{2}$, we have a
capped capillary with spins subject to bulk field $h$ exhibiting 
continuous CC when $h\equiv h_{co}(L)\approx-\sigma\cos\theta/(m_0 L)$. 
In the limit $L\to 0$ we recover a  wedge, which at bulk two-phase 
coexistence exhibits a 2d filling transition
\cite{ParryRasconWoodPRL00,ParryGreenallWoodJPhysCond2002,AbrahamParryWoodEPL2002,AbrahamMaciolekPRL} when 
$\theta(T_{fill})=\alpha$, whereby the thickness of the adsorbed layer at the
bottom diverges: $l_w\sim (T_{fill}-T)^{-\beta_w}$. Finally if $\alpha,L\to 0$ 
we recover a planar geometry with a critical transition at temperature 
$T_{wet}$, at which the mean layer thickness diverges:
$l_{\pi}\sim (T_{wet}-T)^{-\beta^s}$, and $\theta\to
0$ \cite{Reviews}.\\  
{\indent} The trough exhibits a generalised filling transition which can
be studied using the effective Hamiltonian 
\begin{equation}
H_{TR}[\ell]=\int dx\Big\{ \sigma\sqrt{1+\Big(\frac{d\ell}{dx}\Big
)^2}+W_{tr}(\ell,x;h,\alpha)\Big\}
\label{H3}
\end{equation}
  The binding potential for the
trough, $W_{tr}$, has a contact (attractive) interaction at the wall and a term
$2m_0h\ell(x)$ for $|x|<\frac{L}{2}$ which models the bulk-like field acting on
spins in the strip vertically above the bottom. The model is amenable to transfer matrix
analysis similar to that used for filling in acute wedges
\cite{AbrahamParryWoodEPL2002}. In particular, one can establish that, for all $\alpha\le \pi/2$, close to the 
transition
 the probability distribution function (PDF) for the mid-point ($x=0$) 
interfacial height has the simple scaling form 
$P_{tr}(\ell)\propto e^{-\ell/{\langle\ell\rangle}}$
where the mean-height is given by
\begin{equation}
\langle\ell\rangle=\frac{k_B T}{2(\sigma \sin(\theta-\alpha)-m_0 hL)}
\label{midpointheight}
\end{equation}
This result identifies correctly the phase boundaries for CC, filling and wetting in
the limits discussed above. When $L=0$, we recover the covariance 
between the PDFs at 2d
wedge filling and planar critical wetting written, in an obvious notation,
$P_w(\ell;\theta,\alpha)=P_{\pi}(\ell;\theta-\alpha)$
\cite{ParryGreenallWoodJPhysCond2002}. The covariance of the
PDF therefore
extends to 2d CC and establishes a dimensional
reduction between 2d critical- and 1d complete-wetting for thermal 
systems. This has a number of consequences. First, the 2d exponents satisfy
$\beta_A=\beta_w=\beta^s=1$ and are identified with the 1d 
complete wetting result $\beta_{co}^s(1^+)$ noted earlier. Second, the 
invariance of the
PDF necessarily implies that the thermal interfacial wandering exponent
$\zeta(d)$ must satisfy $\zeta(2)=1/2$ and $\zeta(1^+)=1$. These last results 
are
not new but they point to a very deep connection between substrate 
geometry and interfacial fluctuations which is not yet fully explored.
\vspace*{.25cm}

\begin{figure}[thb]
\hspace*{-.25cm}
\includegraphics[width=7.5cm,angle=0]{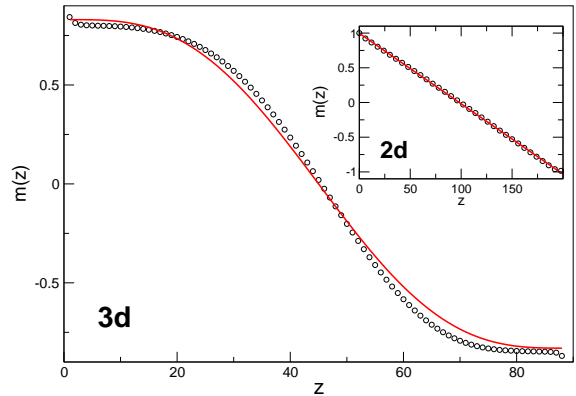}
\caption{Representative Monte Carlo simulation results (points) for the mid-point magnetization profile in 2d and 3d capped capillaries compared with theoretical predictions (\ref{FS2D}), (\ref{FS3D}) (curves). The temperatures, $T=0.741 T^{bulk}_c$ (2d) and $T=0.830 T^{bulk}_c$, respectively, were chosen to avoid the bulk critical region and also the roughening transition (in 3d).}\label{fig3}
\end{figure}

{\indent} Our discussion has focused so far on the critical singularities in a
 semi-infinite capillary. Fluctuation effects and
covariances are also apparent in a capillary of finite-depth, particularly 
if the pressure (or bulk field for Ising systems) is tuned to the
condensation value, $p=p_{co}(L)$, of the infinite capillary. Here we focus on
systems with short-ranged forces  where the geometrical covariances 
are most striking and we can  compare with
simulation studies. We consider only the case of strongly adsorbing walls
($\theta=0$), although the 2d case is insensitive to $\theta$ and
 show that the fluctuations of the meniscus in both 2d and 3d capillaries are 
related, via geometrical
 covariance, to universal FS effects in an entirely different 2d geometry.

 {\indent}Consider an infinitely long, 2d Ising magnet with boundaries at 
$z=0,D$ along
which the spins are subject to fields $h_1$ and $-h_1$ respectively
\cite{ParryEvansPRL,ParryEvansNicolaidesPRL91}. In zero
bulk field, the lower wall is wet by up spins for $T\ge T_{wet}$ while the upper
is wet by down spins. The FS effects are known to fall into two universality 
classes. Exactly at the critical wetting temperature $T=T_{wet}$ the 
profile contains a scaling contribution \cite{ParryEvansNicolaidesPRL91}
\begin{equation}
\frac{m(z)}{m_0}=1-\frac{2z}{D}
\label{FS2D}
\end{equation}
while for $T_c^{bulk}>T>T_{wet}$, 
\begin{equation}
\frac{m(z)}{m_0}=1-\frac{2z}{D}+\frac{1}{\pi}\sin\left(\frac{2\pi z}{D}\right)
\label{FS3D}
\end{equation}
These results are valid in the scaling limits $z,D\to\infty$ with $z/D$
arbitrary and are verified by exact Ising calculations
\cite{AbrahamSvrakicUptonPRL92,SteckiMaciolek}. Both are indicative of
large-scale interfacial fluctuations whereby the roughness scales with the
separation $D$.\\
{\indent}Now consider the
magnetisation profile $m(z)$ measured along the central axis $x=0$ of a finite-depth capped
capillary at bulk field tuned to condensation, {\it i.e.} $h=h_{co}(L)$. If the capillary is
2d, covariance with 2d critical wetting implies we should expect 
(\ref{FS2D}). In a 3d capillary on the other
hand, transfer matrix analysis of (\ref{H2}), with short-ranged forces, leads us to 
the prediction (\ref{FS3D}).

{\indent} In order to test these predictions Monte Carlo simulations were performed
 using a Metropolis algorithm 
for system sizes $L=13$, $D=200$ (2d) and $L=11$, $D=88$, $M=9000$ (3d). The length $M$ in
the direction along the 3d capillary is sufficiently large to preclude FS.
More quantitatively, transfer matrix analysis shows these are negligible if
$\exp{\left(-\frac{6\pi^3\omega M}{L(\kappa D)^2}\right)}<<1$ where $\omega $ is the dimensionless wetting
parameter \cite{Reviews} and $\kappa$ the inverse bulk correlation length. We mimic the
capillary cap geometry by fixing all spins on all surfaces to $+1$ except along the
top line/plane $z=D$ where they are fixed to $-1$. Periodic boundary
conditions apply (in 3d) along the capillary. First we determined the CC line 
$h_{co}(L)$ for the 2d and 3d open capillaries, where periodic boundary
conditions apply at the top ($z=L$) and bottom ($z=0$), using standard 
multicanonical and histogram reweighting techniques. The measured form of the 
axial magnetisation profiles at $h_{co}$ are shown in Fig.\ \ref{fig3}. The comparison with the
theoretical prediction is good in both dimensions but particularly so 
in 2d confirming the covariance with critical wetting. Residual
discrepancies in the 3d case are attributable to a failure to fully
attain the scaling limit $D\to\infty$ which, in view of the requirement
$M>D^2$, is computationally intractable.

{\indent} We have shown that continuous CC is intimately related to a
number of other surface phase transitions and highlights the deep
connection between interfacial behaviour, fluctuation effects and substrate
geometry. Our predictions of distinct critical singularities for adsorption and
desorption in deep capillaries should be verifiable in experiments similar to
those reported in (\cite{MisturaPRL02}). Finally we remark that the present
discussion has been largely limited (in 3d) to the most experimentally
accessible case where confining walls are completely wet ($\theta=0$). The phenomenology will 
be significantly enriched if one considers walls with non-zero contact angle since the
unbinding of the meniscus at CC for adsorption and/or desorption may become 
first-order. This intriguing possibility requires further study.

\acknowledgments{AOP thanks the Universidad Carlos III de Madrid for a short-stay grant (2006). CR acknowledges support from project MOSSNOHO (S-0505/ESP/000299) from Comunidad Aut\'{o}noma de Madrid and by the MEC through grant MOSAICO.}

\end{document}